\def\etal{{\it et al.}}

\def\~{{$\tilde{\phantom{a}}$}}

\def\mbar{{\overline m}}

\documentclass [12pt] {article}
\usepackage{epsfig}
\usepackage{color}

\newcommand{\postscript}[2]
 {\setlength{\epsfxsize}{#2\hsize}
  \centerline{\epsfbox{#1}}}

\renewcommand{\arraystretch}{1.5}

\def\thebibliography#1{\section{References}\markboth
 {REFERENCES}{REFERENCES}\list
 {[\arabic{enumi}]}{\settowidth\labelwidth{[#1]}\leftmargin\labelwidth
 \advance\leftmargin\labelsep
 \usecounter{enumi}}
 \def\newblock{\hskip .11em plus .33em minus -.07em}
 \sloppy
 \sfcode`\.=1000\relax}
\def\upcite#1{\raise6pt\hbox{\scriptsize
\cite{#1}}}
  \def\lsim{\mathrel {\vcenter {\baselineskip 0pt \kern 0pt
    \hbox{$<$} \kern 0pt \hbox{$\sim$} }}}
    \def\gsim{\mathrel {\vcenter {\baselineskip 0pt \kern 0pt
    \hbox{$>$} \kern 0pt \hbox{$\sim$} }}}

\def\r{\\ [2pt]} 
\def\s{\\ [-8pt]}  
\def\hline{\noalign{\hrule \vskip2pt}}

\def\|{\ifmmode\Vert\else \char`\|\fi}
\ifx\oldzeta\undefined                          
  \let\oldzeta=\zeta                            
  \def\zzeta{{\raise 2pt\hbox{$\oldzeta$}}}     
  \let\zeta=\zzeta                              
\fi

\ifx\oldchi\undefined                           
  \let\oldchi=\chi                              
  \def\cchi{{\raise 2pt\hbox{$\oldchi$}}}       
  \let\chi=\cchi                                
\fi

\def\frac#1#2{{#1 \over #2}}

\def\half{\ifinner {\scriptstyle {1 \over 2}}
   \else {1 \over 2} \fi}

\def\simge{\mathrel{%
   \rlap{\raise 0.511ex \hbox{$>$}}{\lower 0.511ex \hbox{$\sim$}}}}
\def\simle{\mathrel{
   \rlap{\raise 0.511ex \hbox{$<$}}{\lower 0.511ex \hbox{$\sim$}}}}

\def\buildchar#1#2#3{{\null\!                   
   \mathop#1\limits^{#2}_{#3}                   
   \!\null}}                                    
\def\overcirc#1{\buildchar{#1}{\circ}{}}

\def\slashchar#1{\setbox0=\hbox{$#1$}           
   \dimen0=\wd0                                 
   \setbox1=\hbox{/} \dimen1=\wd1               
   \ifdim\dimen0>\dimen1                        
      \rlap{\hbox to \dimen0{\hfil/\hfil}}      
      #1                                        
   \else                                        
      \rlap{\hbox to \dimen1{\hfil$#1$\hfil}}   
      /                                         
   \fi}                                         %

\def\subrightarrow#1{
  \setbox0=\hbox{
    $\displaystyle\mathop{}
    \limits_{#1}$}
  \dimen0=\wd0
  \advance \dimen0 by .5em
  \mathrel{
    \mathop{\hbox to \dimen0{\rightarrowfill}}
       \limits_{#1}}}                           

\def\overlay#1#2{\ifmmode%
\setbox0=\hbox{$#1$}%
\setbox1=\hbox to\wd0{\hss$#2$\hss}\else%
\setbox0=\hbox{#1}%
\setbox1=\hbox to\wd0{\hss#2\hss}\fi%
#1\hskip-\wd0\box1 }

\def\pmb#1{\leavevmode\setbox0=\hbox{#1}%
\kern-.02em\copy0\kern-\wd0
\kern.04em\copy0\kern-\wd0
\kern-.02em\raise.04em\box0 }

\def\vereq#1#2{\lower3pt\vbox{\baselineskip1.5pt \lineskip1.5pt
\ialign{$\m@th#1\hfill##\hfil$\crcr#2\crcr\sim\crcr}}}

\def\tensor#1{\protect\@ontopof{#1}{\leftrightarrow}{1.15}\mathord{\box2}}
\def\overstar#1{\protect\@ontopof{#1}{\ast}{1.15}\mathord{\box2}}
\def\overdots#1{\protect\@ontopof{#1}{\cdots}{1.0}\mathord{\box2}}
\def\overcirc#1{\protect\@ontopof{#1}{\circ}{1.2}\mathord{\box2}}
\def\loarrow#1{\protect\@ontopof{#1}{\leftarrow}{1.15}\mathord{\box2}}
\def\roarrow#1{\protect\@ontopof{#1}{\rightarrow}{1.15}\mathord{\box2}}

\def\@ontopof#1#2#3{%
{\mathchoice
{\@@ontopof{#1}{#2}{#3}\displaystyle\scriptstyle}%
{\@@ontopof{#1}{#2}{#3}\textstyle\scriptstyle}%
{\@@ontopof{#1}{#2}{#3}\scriptstyle\scriptscriptstyle}%
{\@@ontopof{#1}{#2}{#3}\scriptscriptstyle\scriptscriptstyle}%
}%
}

\def\@@ontopof#1#2#3#4#5{%
\setbox0=\hbox{$#4#1$}%
\setbox1=\hbox{$#5#2$}%
\setbox2=\hbox{}\ht2=\ht0 \dp2=\dp0 %
\ifdim\wd0>\wd1 %
\setbox1=\hbox to\wd0{\hss\box1\hss}%
\mathord{\rlap{\raise#3\ht0\box1}\box0}%
\else   %
\setbox1=\hbox to.9\wd1{\hss\box1\hss}%
\setbox0=\hbox to\wd1{\hss$#4\relax#1$\hss}%
\mathord{\rlap{\copy0}\raise#3\ht0\box1}%
\fi
}%

\def\lambdabar{\protect\@lambdabar}
\def\@lambdabar{%
\relax
\bgroup
\def\@tempa{\hbox{\raise.73\ht0
\hbox to0pt{\kern.25\wd0\vrule width.5\wd0
height.1pt depth.1pt\hss}\box0}}%
\mathchoice{\setbox0\hbox{$\displaystyle\lambda$}\@tempa}%
{\setbox0\hbox{$\textstyle\lambda$}\@tempa}%
{\setbox0\hbox{$\scriptstyle\lambda$}\@tempa}%
{\setbox0\hbox{$\scriptscriptstyle\lambda$}\@tempa}%
\egroup
}

\def\corresponds{{\lower.2ex\hbox{=}}{\rm\kern-.75em^\triangle}}
\def\succsim{\succ\kern-.9em_\sim\kern.3em}
\def\precsim{\prec\kern-1em_\sim\kern.3em}
\def\slantfrac#1#2{\kern1em^{#1}\kern-.3em/\kern-.1em_{#2}}

\begin{document}                                                                

\renewcommand{\arraystretch}{1.5}

\begin{center}
{\Large\bf The Weizs\"acker-Williams Approximation 
\\

\medskip

to Trident Production in Electron-Photon Collisions}

\bigskip
C.~Bula and K.T.~McDonald
\\
{\sl Joseph Henry Laboratories, Princeton University, Princeton, NJ 08544}
\\
(February 28, 1997)
\bigskip

{\Large\bf Abstract}
\end{center}

The appears to exist no detailed calculation of the multiphoton trident
process $e + n \omega_0 \to e' + e^+e^-$, which can occur during the
interaction of an electron beam with an intense laser beam.  We present
a calculation in the Weizs\"acker-Williams approximation that is in good
agreement with QED calculations for the
weak-field case.  

\section{The Weak-Field Case}

\subsection{Introduction}

As a test of the applicability of the Weizs\"acker-Williams method we first
consider the weak-field case in which only a single initial photon is
involved:
\begin{equation}
\omega_0 + e \to e' + e^+e^-.
\label{eq1}
\end{equation}
A complete calculation for this process for unpolarized electrons and photons
is available \cite{Vortruba}, but apparently
the analytic form is too complex to be enlightening.  A useful summary is
given in sec.~11-4 of ref.~\cite{Jauch}.

Most discussions of reaction (\ref{eq1}) use the frame in which the initial
electron is at rest.  In our recent experiment \cite{Burke,Bamber}
in which trident production is a background, the electron was ultrarelativistic 
in the lab
frame.  To be able to discuss the process in either frame it is useful to
emphasize the relativistic invariants of the problem.  In particular, we
use 
\begin{equation}
s = (\omega_0 + e)^2.
\label{eq2}
\end{equation}
where $s$ is the square of the center of mass energy of reaction (\ref{eq1})
and in eq.~(\ref{eq2})
$\omega_0$ and $e$ represent the 4-momenta of the initial electron and
photon.  The threshold for reaction (\ref{eq1}) is $s_{\rm min} = 9m^2$,
corresponding to the case when all final-state particles are at rest in the
c.m.~frame.

Just above threshold the cross section for the trident process (\ref{eq1}) 
varies as 
\begin{equation}
\sigma_T = 9.2 \times 10^{-4} \alpha r_0^2 \left( {s - 9m^2 \over m^2} 
\right)^2,
\qquad (s - 9m^2 \ll m^2),
\label{eq3}
\end{equation}
where $\alpha = e^2/\hbar c$ is the fine structure constant, $r_0 = e^2/mc^2$
is the classical electron radius, $m$ is the electron rest mass and $c$ is the
speed of light.  Far above threshold the cross section varies as
\begin{equation}
\sigma_T = \alpha r_0^2 \left( {28 \over 9} \ln{s \over m^2} - {100 \over 9} 
\right), \qquad (s  \gg 9m^2),
\label{eq4}
\end{equation}

\begin{figure}[htp]  
\postscript{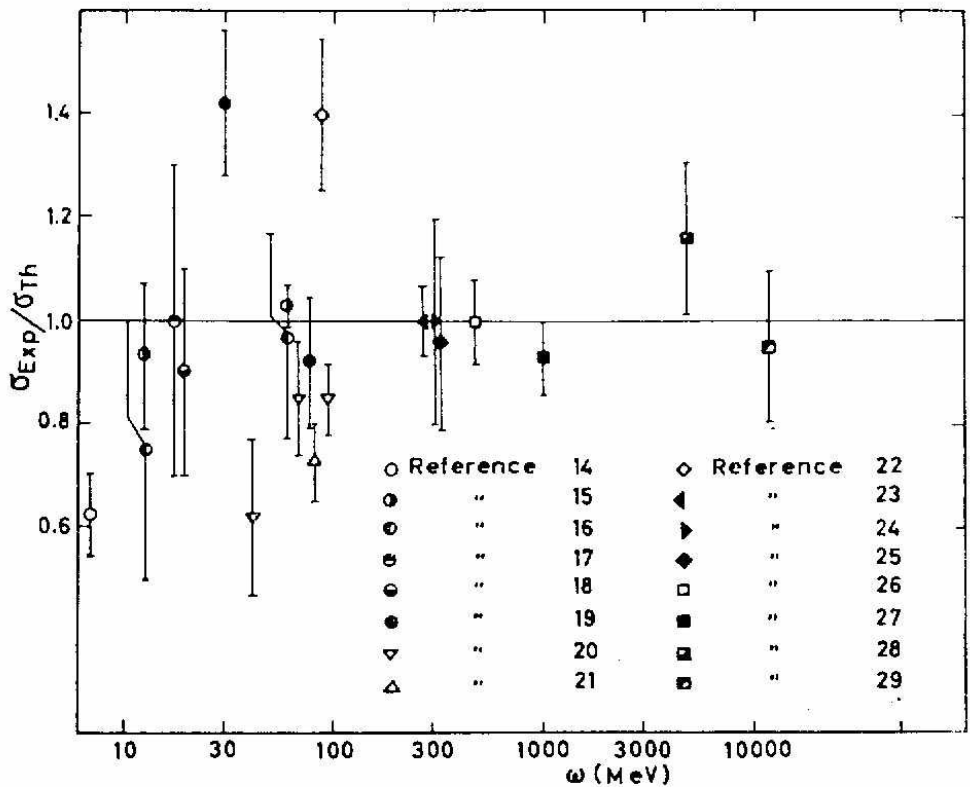}{0.8}
\begin{center}
\parbox{5.5in} 
{\caption[ Short caption for table of contents ]
{\label{morkfig} Comparison of theory and experiment for reaction (\ref{eq1}).
From \cite{Mork}, which includes references to the experiments.
}}
\end{center}
\end{figure}

In our experiment, we were near threshold for trident production, but 
for $s - s_{\rm min}$ extended up to ${\cal O}(m^2)$.  
Hence, we were between the regions of applicability of
the asymptotic relations (\ref{eq3}) and (\ref{eq4}).  A numerical tabulation
of the trident cross section has been given by Mork \cite{Mork}
based on the analytic calculation of Vortruba \cite{Vortruba}.
Figure~\ref{morkfig} compares theory and experiment for reaction (\ref{eq1}).

  We will compare
the Weizs\"acker-Williams approximation to the result of ref.~\cite{Mork} below.

\begin{figure}[htp]  
\postscript{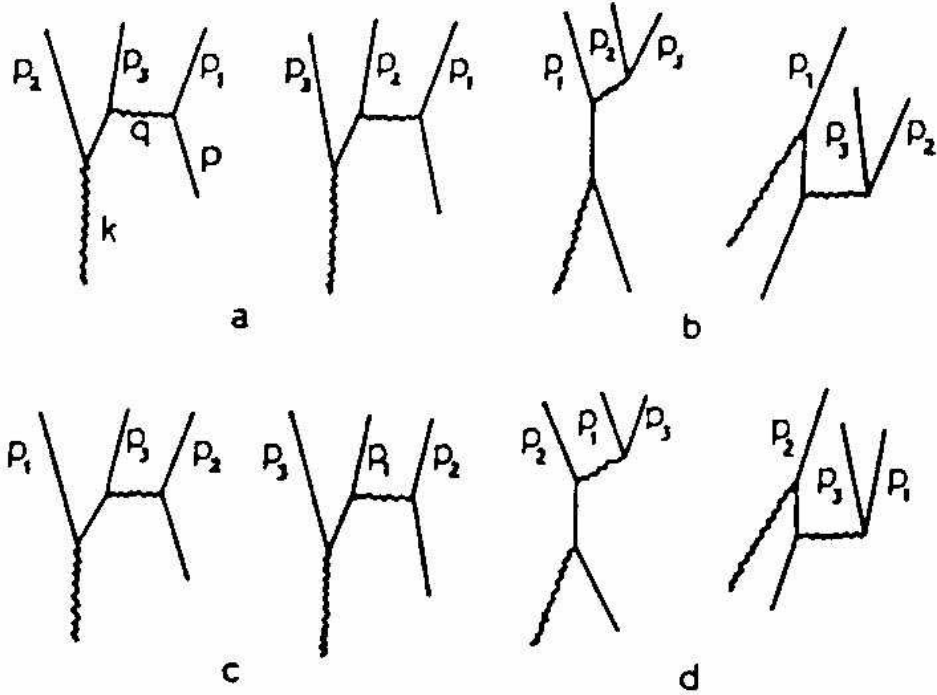}{0.8}
\begin{center}
\parbox{5.5in} 
{\caption[ Short caption for table of contents ]
{\label{morkfig1} Feynman diagrams for reaction (\ref{eq1}).  Diagrams
(b) and (d) are called $\gamma$-$e$ or Compton diagrams.
}}
\end{center}
\end{figure}

Mork also reported numerical results from a simplified calculation by
Borsellino \cite{Borsellino} of
trident production in which  diagrams (b) and (d) of Fig.~\ref{morkfig1}
are neglected.
These diagrams are referred to as $\gamma$-$e$ or as Compton
diagrams in that the initial-state photon couples directly to the initial
electron rather than the $e^+e^-$ pair.  Well above threshold the Compton
diagrams contribute little to the cross section, while near threshold they
interfere to reduce the cross section, as summarized in Fig.~\ref{fig3}
below.  Thus the neglect of the Compton diagrams results in an overestimate
of the cross section.

\subsection{The Weizs\"acker-Williams Approximation}

In a frame in which the initial electron is ultrarelativistic its electric
and magnetic fields are nearly transverse and of nearly equal magnitude.  That 
is, the fields appear to be almost identical to those of a packet of real
photons, which we label by $\omega_1$.  A Fourier transform of the
time dependence of the electron's field integrated over observers at impact
parameters $b > b_{\rm min}$ to the electron's trajectory yields the photon
number spectrum
\begin{equation}
N(\omega_1) \approx {2\alpha \over \pi\omega_1} \ln {\gamma \over \omega_1
b_{\rm min}},
\label{eq5}
\end{equation}
where $\gamma$ is the Lorentz factor for the initial electron \cite{Jackson}.
We then suppose that a (virtual) photon $\omega_1$
from this spectrum combines with the incident photon $\omega_0$ to produce
an $e^+e^-$ pair via the Breit-Wheeler process \cite{Breit}:
\begin{equation}
\omega_0 + \omega_1 \to e^+e^-.
\label{eq6}
\end{equation}
The Weizs\"acker-Williams approximation to the trident cross section is then
\begin{equation}
\sigma_T = {2\alpha \over \pi} \int_{\omega_{1,\rm min}}^{\omega_{1,\rm max}}
{d\omega_1 \over \omega_1} \ln {\gamma \over \omega_1 b_{\rm min}}
\sigma_{\rm BW}(\omega_0,\omega_1),
\label{eq7}
\end{equation}
where $\sigma_{\rm BW}$ is the Breit-Wheeler cross section.

The Breit-Wheeler cross section can be expressed in terms of $s'$, the
square of the center of mass energy of the photon-photon system.  For 
a frame in which the two photons collide head on, $s' = 4\omega_0\omega_1$, 
where in this expression $\omega$ stands for the photon energy.  Then
\begin{equation}
\sigma_{\rm BW} = 4\pi r_0^2 {m^2 \over s'} \left[ {3 - \beta^4 \over 2} 
\ln {1 + \beta \over 1 - \beta} - \beta(2 - \beta^2) \right],
\label{eq8}
\end{equation}
where
\begin{equation}
\beta = \sqrt{1 - {4m^2 \over s'}}
\label{eq8a}
\end{equation}
is $v/c$ of the positron (or partner 
electron) in the pair rest frame.  The threshold condition is, of course,
$s'_{\rm min} = 4m^2$.  The asymptotic forms are
\begin{equation}
\sigma_{\rm BW} \approx \pi r_0^2 \beta, \qquad (\beta \ll 1 \Leftrightarrow
s' - 4m^2 \ll m^2),
\label{eq9}
\end{equation}
and
\begin{equation}
\sigma_{\rm BW} \approx 2\pi r_0^2 {m^2 \over s'} \left( \ln {s' \over m^2} - 1
\right), \qquad (s' \gg 4m^2).
\label{eq10}
\end{equation}
See, for example, sec.~13-3 of ref.~\cite{Jauch}.

Before inserting eq.~(\ref{eq8}) into (\ref{eq7}) the latter should be put
into a form that is more manifestly covariant.
Our approach is to replace $\omega_1$ in (\ref{eq7}) by 
$s' = 4\omega_0\omega_1$.  Immediately $d\omega_1/\omega_1 = ds'/s'$.  

Then the lower limit of
integration, originally $\omega_{1,\rm min}$, becomes $s'_{\rm min} = 4m^2$.  

The
upper limit of integration becomes $s'_{\rm max}$ for the Breit-Wheeler process
embedded in the trident reaction (\ref{eq1}).  Another interpretation of
$s'$ is as the square of the invariant mass of the $e^+e^-$ pair:
$s' = m_{e^+e^-}^2$.  Then $s'_{\rm max}$ occurs when as much energy as possible
goes into the mass of the $e^+e^-$ pair.  This occurs when both the pair and
the scattered electron $e'$ are at rest in the c.m.\ frame of reaction 
(\ref{eq1}).  In this case
\begin{equation}
s = (m + m_{e^+e^-})^2 = (m + \sqrt{s'_{\rm max}})^2, \qquad \mbox{and hence} 
\qquad
s'_{\rm max} = (\sqrt{s} - m)^2,
\label{eq11}
\end{equation}
where $s$ is the square of the c.m.\ energy of reaction (\ref{eq1}).

Finally,
 we need to reinterpret the argument of the logarithm in eq.~(\ref{eq7}).
It should be an invariant, should be greater than 1, and should have $\omega_1$
in the denominator.  The simplest form is then $s'_{\rm max}/s'$.  This could be
multiplied by a number of order 1, which is the usual ambiguity of the
Weizs\"acker-Williams method.

Altogether, the proposed invariant combination of (\ref{eq7}) and (\ref{eq8}) is
\begin{eqnarray}
\sigma_T & = & {2\alpha \over \pi} \int_{4m^2}^{s'_{\rm max}}
{ds' \over s'} \ln {s'_{\rm max} \over s'} \sigma_{\rm BW}(s') \nonumber \\
& = & 8\alpha r_0^2 \int_{4m^2}^{s'_{\rm max}}
{m^2 ds' \over s^{'2}} \ln {s'_{\rm max} \over s'} \left[ {3 - \beta^4 \over 2} 
\ln {1 + \beta \over 1 - \beta} - \beta(2 - \beta^2) \right],
\label{eq12}
\end{eqnarray}
where $\beta$ and $s'_{\rm max}$ 
are given by eqs.~(\ref{eq8a}) and (\ref{eq11}).
This form can be evaluated in a frame in which the initial electron is at
rest even though it is unclear that the field of the electron is equivalent
to a collection of real photons in this frame.

\begin{figure}[htp]  
\postscript{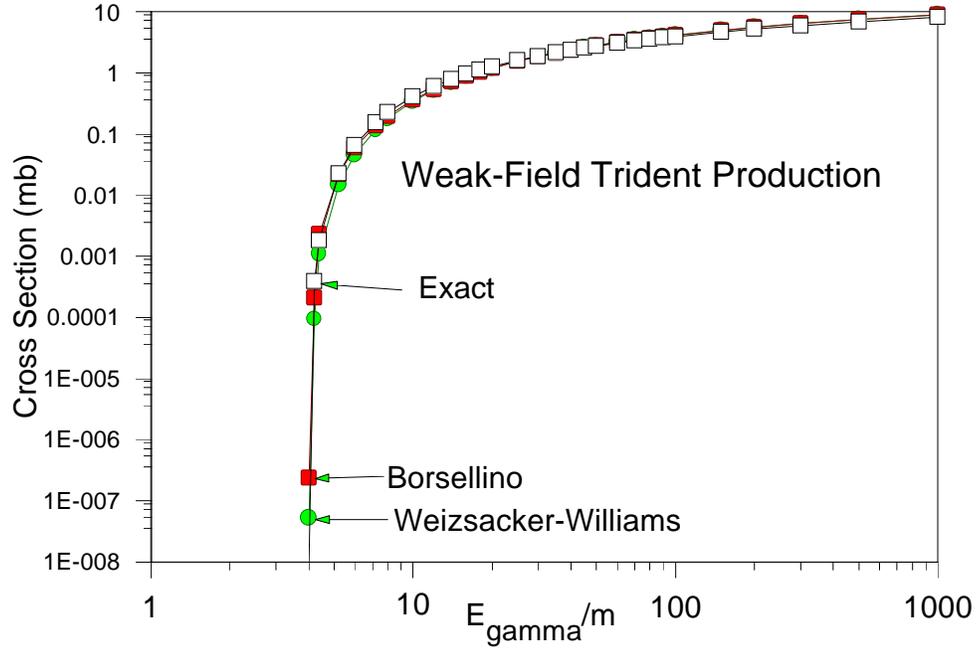}{0.8}
\begin{center}
\parbox{5.5in} 
{\caption[ Short caption for table of contents ]
{\label{fig3} Calculated cross sections for the trident process (\ref{eq1})
for the initial electron at rest and initial photon of energy $E_{\gamma}$.
Then $E_{\gamma}/m = (s - s_{\rm min})/2m^2$, the invariant measure of the 
energy
of the interaction above threshold.
}}
\end{center}
\end{figure}

\begin{figure}[htp]  
\postscript{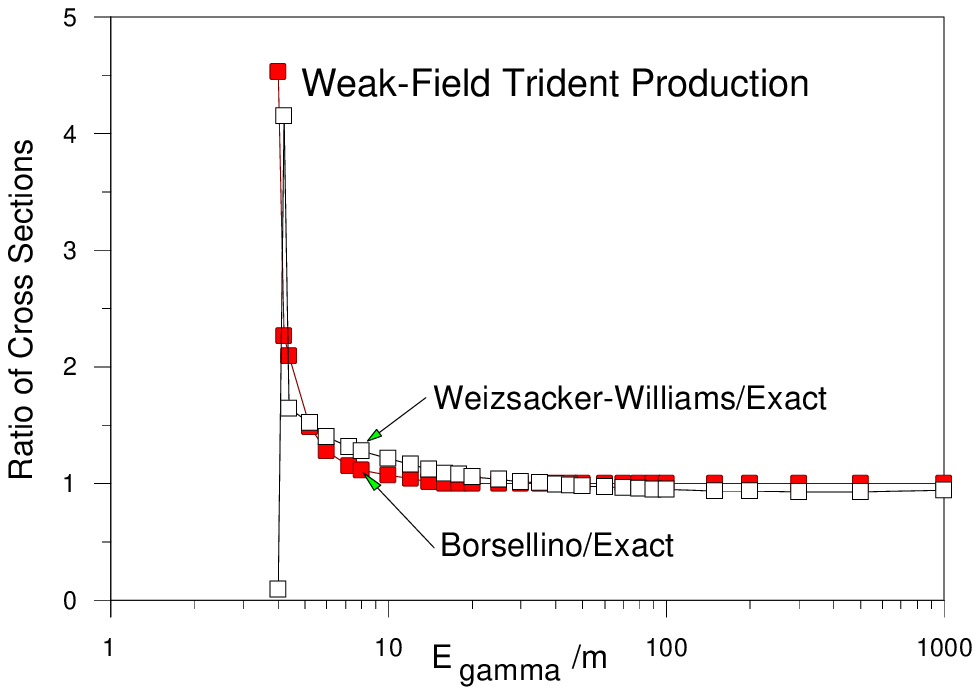}{0.8}
\begin{center}
\parbox{5.5in} 
{\caption[ Short caption for table of contents ]
{\label{fig4} Ratio of calculated cross sections of the trident process 
(\ref{eq1}).
}}
\end{center}
\end{figure}

\begin{figure}[htp]  
\postscript{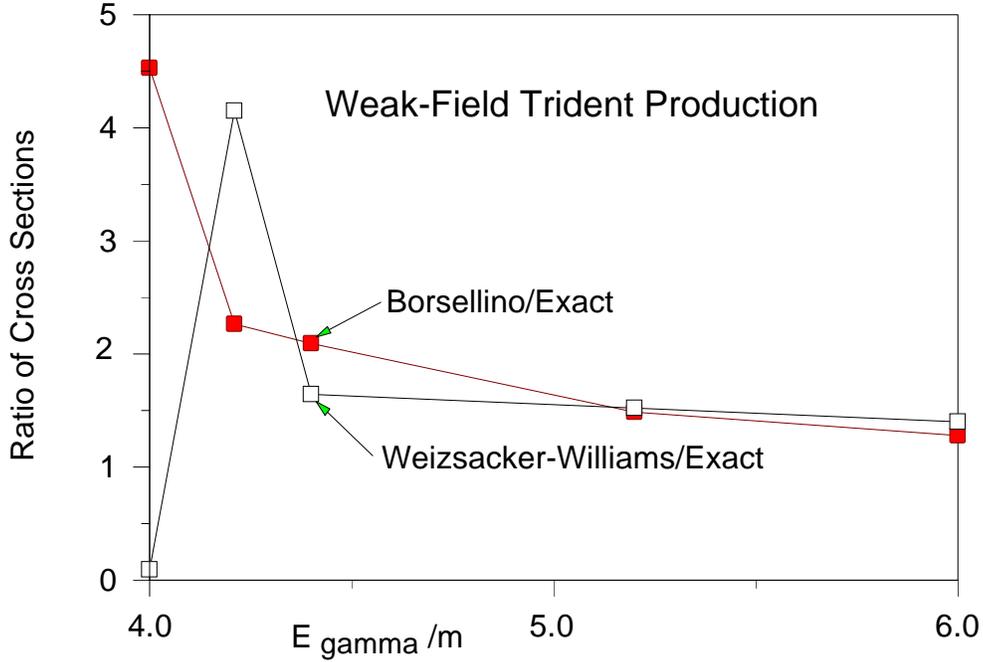}{0.8}
\begin{center}
\parbox{5.5in} 
{\caption[ Short caption for table of contents ]
{\label{fig5} Ratio of calculated cross sections of the trident process 
(\ref{eq1}).
}}
\end{center}
\end{figure}

Figures \ref{fig3}-\ref{fig5} show results of numerical calculations of
eq.~(\ref{eq12}) along with the ``exact'' cross section as tabulated by Mork
and the cross sections calculated by Borsellino by ignoring diagrams (b) and
(d) of Fig.~\ref{morkfig1}.  The agreement of the Weizs\"acker-Williams
approximation and the exact calculation is fairly good, although worst near
threshold where the rate is very low.

The results are plotted as a function of the invariant $(s - s_{\rm min})/2m^2$
which is a measure of how far the reaction is above threshold.  For the
initial electron at rest this invariant is $E_{\gamma}/m$ where $E_{\gamma}$ is
the energy of the initial photon.

In the Weizs\"acker-Williams approximation the initial photon $\omega_0$
interacts only with the $e^+e^-$ pair, not directly with the initial
electron.  Thus the approximation neglects diagrams (b) and (d) of
Fig.~\ref{morkfig1}.  This feature is shared with the approximation of
Borsellino \cite{Borsellino}, 
and indeed  Figs.~\ref{fig3}-\ref{fig5} show that the
Weizs\"acker-Williams approximation tracks Borsellino's results more 
closely than the ``exact'' results.

The Weizs\"acker-Williams approximation corresponds to the use of transverse
but otherwise unpolarized virtual photons.  This feature is shared with
the exact calculation of diagrams (a) and (c) of Fig.~\ref{morkfig1}.
Only for the neglected diagrams (b) and (d) could there be polarization
of the virtual photons in case the initial photon is polarized.   For the
Compton diagrams (b) and (d) the virtual photons would take on the
polarization of the initial photon for energies of the virtual photon
near the maximum.


\section{The Strong-Field Case}

Multiphoton trident production can occur in a strong field of initial-state 
photons:
\begin{equation}
e + n\omega_0 \to e' + e^+e^-.
\label{eq13}
\end{equation}
Extrapolating from eq.~(\ref{eq12}), we propose the Weizs\"acker-Williams
approximation for reaction (\ref{eq13}) be formulated as
\begin{equation}
Rate_T  = {2\alpha \over \pi} \sum_n \int_{4\mbar^2}^{s'_{n,\rm max}}
{ds'_n \over s'_n} \ln {s'_{n,\rm max} \over s'_n} Rate_{\rm BW}(s'_n,\eta),
\label{eq14}
\end{equation}
where $\mbar = m\sqrt{1 + \eta^2}$ is the shifted mass of the electron in
the strong field and $\eta = e{\cal E}_{\rm rms}/m\omega c$
is the field-strength parameter.
In this we calculate a rate rather than a cross section, using the results
of Nikishov and Ritus \cite{Nikishov}.  
For $n$ initial-state (laser) photons, the sub-process
is the multiphoton Breit-Wheeler reaction
\begin{equation}
n\omega_0 + \omega_1 \to e^+e^-,
\label{eq14a}
\end{equation}
for which the square of the c.m.~energy is
\begin{equation}
s'_n = (n\omega_0 + \omega_1)^2,
\label{eq15}
\end{equation}
where in this expression $\omega$ stands for the 4-momentum of a photon.
Equation (\ref{eq11}) becomes
\begin{equation}
s'_{n,\rm max} = (\sqrt{s_n} - \mbar)^2, \qquad \mbox{with} 
\qquad 
s_n = (e + n\omega_0)^2 = \mbar^2 + 2n(e \cdot \omega_0),
\label{eq16}
\end{equation}
where $e$ and $\omega_0$ are the initial-state 4-momenta (including mass-shift
effects, hence $e^2 = \mbar^2$).  

If $s'_{n,\rm max} < 4\mbar^2$ there is no contribution at order $n$.  This
condition can be stated another way.  The threshold condition is that
the final-state electron and the $e^+e^-$ pair are both at rest in the
c.m.~frame of the $e + n\omega_0$ system, and that $m_{e^+e^-} = 2\mbar$.
That is,
\begin{equation}
s_{n,\rm min} = (e + n\omega_0)^2 \ge (\mbar + 2\mbar)^2 = 9\mbar^2.
\label{eq17}
\end{equation}
For a head-on collision between a relativistic electron and the initial-state
photons this becomes
\begin{equation}
\mbar^2 + 4nE_0\omega_0 \ge 9\mbar^2, \qquad \mbox{or} \qquad
n \ge {2\mbar^2 \over E_0\omega_0},
\label{eq18}
\end{equation}
where $E_0$ is the energy of the initial electron.  Strictly speaking,
$E_0 = q_0$, the quasienergy of the initial electron which is related by
\begin{equation}
q = e + {\eta^2m^2 \over 2(e \cdot \omega_0)} \omega_0,
\label{eq19}
\end{equation}
where $e$, $q$ and $\omega_0$ are 4-momenta in this expression.  For
our example, $q_0 \approx E_0 + \eta^2m^2/4E_0 \approx E_0$.

For $E_0$ = 46.6 GeV and $\omega_0$ = 2.3 eV we must have $n \ge 5$.
As noted in sec.~1.2, this corresponds to all initial-state photons
coupling only to the $e^+e^-$ pair, and ignores the strong-field generalizations
of diagrams (b) and (d) of Fig.~\ref{morkfig1}.

In the Weizs\"acker-Williams approximation the virtual photon $\omega_1$
is unpolarized even if the initial-state photon $\omega_0$ is polarized.
Hence the Breit-Wheeler rate used in eq.~(\ref{eq14}) should be
for unpolarized $\omega_1$ but with whatever polarization holds for the
initial-state photons $\omega_0$.

\subsection{Numerical Results}

The above procedures have been implemented in a numerical simulation
\cite{Bamber}.  

The requirements (\ref{eq16}) and (\ref{eq17}) that energy be conserved during 
pair creation has a striking effect on the calculated rate.  First, the
minimum number of laser photons is $n = 5$ (and it turns out that there is
no significant rate unless $n > 5)$.  Second, the maximum energy of the
virtual photon, $\omega_{1,\rm max}$, that can contribute is much less than
the electron beam energy $E_0$.  The latter point can be anticipated by
the approximation of head-on collisions for which eq.~(\ref{eq16}) tells us
\begin{equation}
\omega_{1,\rm max} = E_0 - {\mbar^2 \over 2n\omega_0} \left( \sqrt{1 +
{4nE_0\omega_0 \over \mbar^2}} - 1 \right).
\label{eq20}
\end{equation}
Some representative values are given in the Table.

\begin{table}[htbp] 
\begin{center}
\parbox{5.5in}  
{\caption[ Short caption for the List of Tables. ]
{\label{table1} Maximum energy $\omega_{1,\rm max}$
of a virtual photon that can contribute to
the trident process for $E_0 = 46.6$ GeV, $\omega_0 = 2.35$ eV and $\mbar = m$
as a function of the number $n$ of laser photons, assuming head-on collisions.
}}
\vskip6pt
\begin{tabular}{cccccc}
\hline\hline
$n$ & 6 & 8 & 10 & 12 & 20 \s
 $\omega_{1,\rm max}$ (GeV) & 25.1 & 27.2 & 28.7 & 30.0 &33.1 \r
\hline\hline
\end{tabular}
\end{center}
\end{table}

\begin{figure}[htp]  
\postscript{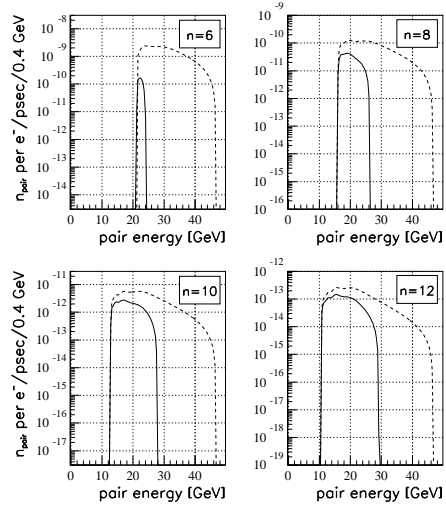}{0.5}
\begin{center}
\parbox{5.5in} 
{\caption[ Short caption for table of contents ]
{\label{fig6} Calculated rates for the trident process (\ref{eq14a}) in 
typical conditions of our experiment \cite{Bamber}.  
Solid curves: $\omega_{1,\rm max}$ taken from
eq.~(\ref{eq16}); dashed curves: $\omega_{1,\rm max} = E_0$.
}}
\end{center}
\end{figure}

Figure \ref{fig6} shows results of the trident-rate calculation
for various numbers of laser photons.  The solid curves are the proper results
while the dashed curves show the effect of setting $\omega_{1,\rm max}$ to 
$E_0$.

\begin{figure}[htp]  
\postscript{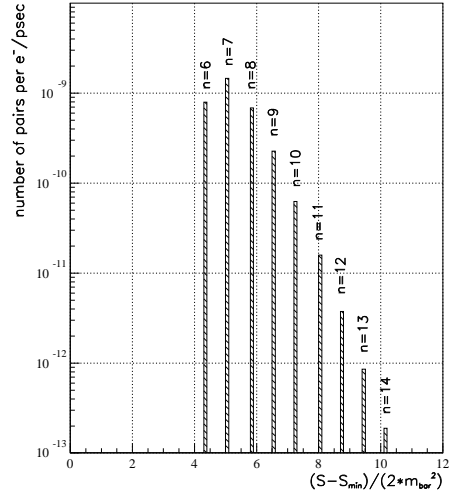}{0.5}
\begin{center}
\parbox{5.5in} 
{\caption[ Short caption for table of contents ]
{\label{fig7} Contributions to the rate of trident production as a function
of the invariant measure $(s - s_{\rm min})/2\mbar^2$ of energy above threshold.
}}
\end{center}
\end{figure}

Figure \ref{fig7} shows the contribution to the rate as a function of the
invariant measure of energy above threshold.  Only the case $n = 6$ is
close enough to threshold that the Weizs\"acker-Williams approximation
is significantly in error, and the sign of the error is to overestimate
the rate.

\begin{figure}[htp]  
\postscript{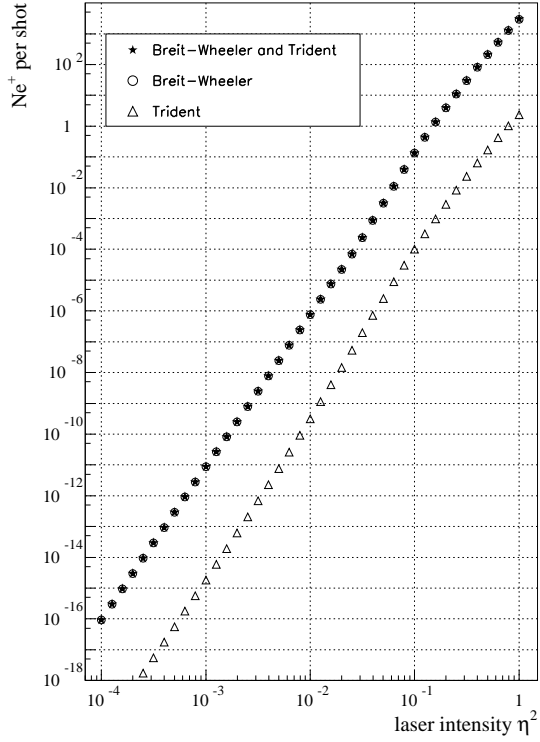}{0.7}
\begin{center}
\parbox{5.5in} 
{\caption[ Short caption for table of contents ]
{\label{fig8} Calculated rates for trident production (\ref{eq14a}) and
two-step pair creation (\ref{eq21}) for typical conditions of our experiment
\cite{Bamber}.
}}
\end{center}
\end{figure}

Finally, Figure \ref{fig8} shows the total rate of trident production 
(\ref{eq14a}) as a function of laser intensity parameter $\eta^2$ for
typical conditions of our experiment \cite{Bamber}.  
Also shown is the calculated rate for pair creation by the two-step process
\begin{equation}
e + n\omega_0 \to e' + \omega_1, \qquad \mbox{followed by} \qquad
\omega_1 + m\omega_0 \to e^+e^-.
\label{eq21}
\end{equation}
The trident process is only a 1\% correction to the two-step production
process.

\end{document}